\documentclass[sigconf,letterpaper,authorversion]{acmart}

\usepackage[english]{babel}
\usepackage{blindtext}

\usepackage{tabularx}
\usepackage[inline]{enumitem}
\usepackage{xurl}

\acmYear{2019}\copyrightyear{2019}
\setcopyright{acmlicensed}
\acmConference[NetAI '19]{NetAI '19: ACM SIGCOMM 2019 Workshop on Network Meets AI \& ML}{August 23, 2019}{Beijing, China}
\acmBooktitle{NetAI '19: ACM SIGCOMM 2019 Workshop on Network Meets AI \& ML, August 23, 2019, Beijing, China}
\acmPrice{15.00}
\acmDOI{10.1145/3341216.3342211}
\acmISBN{978-1-4503-6872-8/19/08}

\usepackage{subcaption}
\usepackage{multirow}
\usepackage{tikz}

\newcommand{\ie}{i.e.,}
\newcommand{\eg}{e.g.,}
\newcommand{\etal}{et~al\@ifnextchar.{}{.\@}}
\newcommand{\etc}{etc\@ifnextchar.{}{.\@}}

\newcommand{\wrt}{w.r.t.\@}

\newcommand{\sref}[1]{Section~\ref{#1}}
\newcommand{\fig}[1]{Figure~\ref{#1}}

\newcommand{\afblock}[1]{\noindent{\textbf{#1}}}

\newcommand{\mbps}{Mbps}

\begin{document}
\newcommand{\name}{D\lowercase{ee}PCCI}
\hyphenation{DeePCCI}

\title[\name: Deep Learning-based\\ Passive Congestion Control Identification]{\name: Deep Learning-based\\ Passive Congestion Control Identification}

\author{Constantin Sander}
\affiliation{%
	\institution{RWTH Aachen University}
}
\email{sander@comsys.rwth-aachen.de}

\author{Jan R\"uth}
\affiliation{%
	\institution{RWTH Aachen University}
}
\email{rueth@comsys.rwth-aachen.de}

\author{Oliver Hohlfeld}
\affiliation{%
	\institution{Brandenburg University of Technology}
}
\email{oliver.hohlfeld@b-tu.de}

\author{Klaus Wehrle}
\affiliation{%
	\institution{RWTH Aachen University}
}
\email{wehrle@comsys.rwth-aachen.de}

\renewcommand{\shortauthors}{Sander and R\"uth, et al.}

\begin{abstract}

Transport protocols use congestion control to avoid overloading a network.
Nowadays, different congestion control variants exist that influence performance.
Studying their use is thus relevant, but it is hard to identify which variant is used.
While passive identification approaches exist, these require detailed domain knowledge and often also rely on outdated assumptions about how congestion control operates and what data is accessible.
We present \name{}, a passive, deep learning-based congestion control identification approach which does not need any domain knowledge other than training traffic of a congestion control variant.
By only using packet arrival data, it is also directly applicable to encrypted (transport header) traffic.
\name{} is therefore more easily extendable and can also be used with QUIC.

 \end{abstract}



\maketitle

\section{Introduction}
Congestion control (CC)~\cite{Jacobson} is a fundamental building block in today's transport protocols and strongly influences the performance of data transmissions.
Initially built in the 1980s to counteract the congestion collapse of the early Internet~\cite{RFC0896}, CC still evolves and new variants, such as BBR~\cite{Cardwell:BBR} or Vivace~\cite{vivace}, emerge.

CC introduces a congestion window (cwnd) that limits the number of unacknowledged bytes in flight.
Each CC algorithm defines how it governs the evolution of the cwnd subject to algorithm-specific congestion signals.
Given the number of CC approaches and their (inter-) performance implications~\cite{Hock:BBRfairness}, it is hence relevant to study the CC usage.
E.g., it is easier to tune a new CC for fairness, if it is known which other algorithms it typically competes with.

However, existing works (\eg{}~\cite{Yang, Oshio, Casagrande:tcpmoon}) for identifying CC variants are not adapted to the latest CC and transport protocol developments.
Extending and maintaining these approaches is complex since it requires detailed domain knowledge to know how CC parameterization and configuration affect their behavior.
This becomes even more critical when CC leaves the kernel and is introduced in user space protocols such as QUIC~\cite{quic} that are comparably easy to change and see already large-scale deployment~\cite{quicWildPAM2018}.
Moreover, many identification approaches are based on fragile assumptions.
For example, they fail when using TCP pacing, \eg{} in combination with RENO~\cite{Jacobson} or CUBIC~\cite{Ha:CUBIC}.
Aggravatingly, all passive approaches known to us are based on the assumption of parsable header information. 
A fully encrypted transport, as for example QUIC implements, renders these designs invalid and would require significant changes if possible at all.
Thus, it is currently challenging to reason about the CC deployment.
As a first step to tackle these challenges, this paper presents \name{}, a supervised deep learning-based approach for passive congestion control identification.
It identifies CC variants solely based on flow packet arrival time information and thus even works on encrypted transport headers.
Moreover, it uses deep learning to learn features --- thereby avoiding manual, domain-specific feature-engineering.
Thus, unlike related approaches, \name{} makes no assumptions other than the availability of flow packet timings nor requires any domain knowledge other than being able to gather training traffic of a CC variant.
We argue that this assumption and hand-tuning free method allows for generic and extensible CC identification in Internet traffic.

Specifically, we present \name{}'s design, its evaluation, and its limitations and make the following contributions:
\vspace{-0.5em}
\begin{itemize}[noitemsep,topsep=5pt,leftmargin=9pt]
	\item We describe the preprocessing of traffic and the deep learning model for identification of congestion control variants.
	\item We present how to train the model for multiple congestion control variants with labeled data generated in a testbed.
	\item We evaluate the performance in a testbed for CUBIC, RENO, and BBR as major congestion control variants.
	We show that the approach is able to identify flow congestion control variants in various scenarios, but we also present and discuss scenarios where it is unable to identify congestion control variants.
\end{itemize}
\vspace{-.5em}

\afblock{Structure.}
\sref{sec:rw} discusses the state-of-the-art in CC identification and its shortcomings.
\sref{sec:design} describes the design of \name{} while \sref{sec:eval} shows how we generate training data and evaluate our approach.
Finally, \sref{sec:conclusion} concludes the paper and discusses future work.
Our paper does not raise any ethical issues.
\section{Related Work}
\label{sec:rw}
Various works deal with identifying CC variants.
There are two main categories for these approaches: Identification approaches using passive~\cite{Jaiswal:tcpflows, Paxson:tcpanaly, Casagrande:tcpmoon, Hagos:lstm, Oshio} or active~\cite{Yang,Padhye:TBIT} measurements.

\afblock{Active approaches} stimulate CC reactions for detection by actively opening connections and manipulating them.
Padhye and Floyd~propose TBIT~\cite{Padhye:TBIT} which sends crafted TCP segments to web servers to actively trigger congestion control.
It records which segments are sent in reaction to a lost packet, as this reaction differs drastically between the CC variants distinguished in TBIT.

Yang et al.~present CAAI~\cite{Yang} extending TBIT's approach.
CAAI artificially delays ACKs observing all in-flight segments in order to estimate the cwnd of the sender.
CAAI then induces packet loss and extracts characteristic features from the changing cwnd.
These features are subsequently used for classification using random forests.
While both approaches achieve high identification accuracies, relying on active measurements can easily introduce a measurement bias due to wrongly selected hosts.

\afblock{Passive approaches} (as ours) do not interact with hosts and rather rely on traffic traces to infer the used CC variants of traffic flows.
Therefore, they allow gathering information on real traffic which depends on vantage points, not actively selected hosts.

Paxson et al.~and Jaiswal et al.~rebuild TCP state machines with tcpanaly~\cite{Paxson:tcpanaly} and tcpflows~\cite{Jaiswal:tcpflows} to compare received with expected packets.
Both approaches require very detailed CC and even implementation knowledge to rebuild the state machines.
Our approach differs in that it does not need detailed CC knowledge.

\hyphenation{TCPMoon}
Casagrande et al.~\cite{Casagrande:tcpmoon} use characteristic changes in a cwnd as features in their approach TCPMoon.
Different handcrafted rules are checked against these features to distinguish  CCs.
For cwnd estimation, the authors use an RTT estimation based on TCP timestamps.
Identifying flows without the TCP timestamp option or when transport headers are encrypted is hence not feasible with TCPMoon.
As our approach does only observe the behavior of packet arrivals, it does not need any cleartext transport protocol fields.

Oshio et al.~\cite{Oshio} propose a clustering-based method.
They extract features of a cwnd based on an RTT estimate and cluster these to discriminate two competing CC variants.
Our approach differs in that it is not limited to only two competing variants.

Hagos et al.~\cite{Hagos:lstm} use the outstanding bytes between sender and receiver as a rough and noisy cwnd estimate.
This estimate is refined using a recurrent neural network. Sudden decreases in the refined cwnd are used as an estimate of the multiplicative decrease factor which differs between CUBIC, BIC and RENO.
While this approach is similar in that it uses deep learning, it still requires the manually engineered multiplicative decrease factor
and thus only identifies loss-based CC.
Our approach uses an end-to-end deep learning model, identifies also delay-based CC and avoids manual features.
\section{\name{} Design}
\label{sec:design}
In this section, we present the architecture of \name{}.
We show our design goals and list the steps necessary to identify CC variants without variant-specific domain knowledge.

\subsection{Design Goals}
\name{} is designed to allow passive CC variant identification.
It uses supervised deep learning to learn features of CC variant behavior from generated traffic flows and subsequently allows to identify / classify the used CC variants of new flows with these features.
Our design is guided by the following two main points:
\vspace{-.5em}
\begin{itemize}[noitemsep,topsep=5pt,leftmargin=9pt]
	\item First, we want to avoid assumptions such as the availability of headers or specific CC behavior.
	Such assumptions are prone to change or break with slight changes to the algorithms and may render the approach inapplicable.
	We, therefore, do not analyze packet headers, as the assumption of unencrypted packet headers will not hold in the advent of QUIC.
	Moreover, we do not need the ACK control-flow from receiver to sender, as it might be unavailable due to routing asymmetry and encryption.
	\item Second, we want to manually engineer as few features as possible.
	Otherwise, new CC variants would require new manually crafted domain-specific features, so having the ability to automatically learn which features to extract for classification will enable future applicability and easier adaption in case of changes.
\end{itemize}
\vspace{-0.5em}
We argue that these guidelines enable our approach to be robust against minor changes in the Internet and to be easily extendable for major changes by retraining on new data.

\subsection{CC Identification \& Learning Approach}
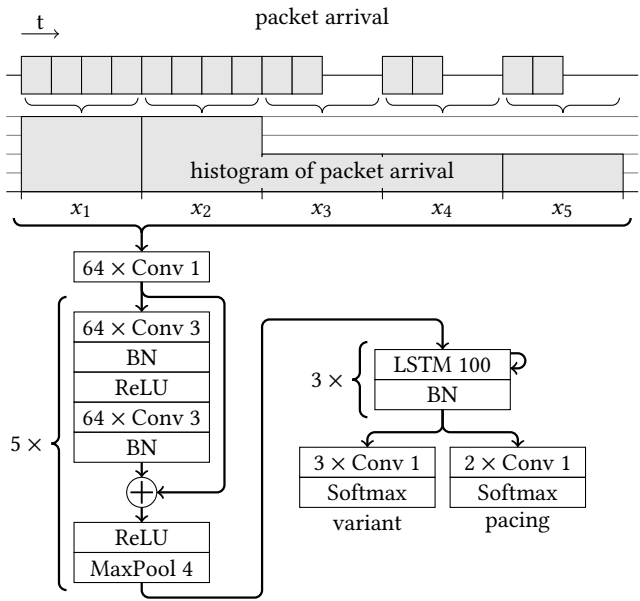
\begin{figure}
	\centering
	\def\svgwidth{\columnwidth}
\begin{tikzpicture}
	\draw (-0.2, 0.25) -- (8.2, 0.25);
	
	\draw[->] (0.0, 0.8) -- (0.5, 0.8);
	\draw (0.25, 0.95) node {t};
	
	\draw (4.0, 1) node {packet arrival};
	
	\fill[draw=black,fill=black!10!] (0,0) rectangle (0.4,0.5);
	\fill[draw=black,fill=black!10!] (0.4,0) rectangle (0.8,0.5);
	\fill[draw=black,fill=black!10!] (0.8,0) rectangle (1.2,0.5);
	\fill[draw=black,fill=black!10!] (1.2,0) rectangle (1.6,0.5);
	
	\fill[draw=black,fill=black!10!] (1.6,0) rectangle (2.0,0.5);
	\fill[draw=black,fill=black!10!] (2.0,0) rectangle (2.4,0.5);
	\fill[draw=black,fill=black!10!] (2.4,0) rectangle (2.8,0.5);
	\fill[draw=black,fill=black!10!] (2.8,0) rectangle (3.2,0.5);
	
	\fill[draw=black,fill=black!10!] (3.2,0) rectangle (3.6,0.5);
	\fill[draw=black,fill=black!10!] (3.6,0) rectangle (4.0,0.5);
	
	\fill[draw=black,fill=black!10!] (4.8,0) rectangle (5.2,0.5);
	\fill[draw=black,fill=black!10!] (5.2,0) rectangle (5.6,0.5);
	
	\fill[draw=black,fill=black!10!] (6.4,0) rectangle (6.8,0.5);
	\fill[draw=black,fill=black!10!] (6.8,0) rectangle (7.2,0.5);
	
	\draw (0.05,-0.05) to[out=270,in=180] +(0.1, -0.1) -- +(0.55, 0) to[out=0,in=90] +(0.65, -0.1)
	+(0.0, 0.0) to[out=90,in=180] +(0.1, 0.1) -- +(0.55, 0.0) to[out=0,in=270] +(0.65, 0.1);
	\draw (1.65,-0.05) to[out=270,in=180] +(0.1, -0.1) -- +(0.55, 0) to[out=0,in=90] +(0.65, -0.1)
	+(0.0, 0.0) to[out=90,in=180] +(0.1, 0.1) -- +(0.55, 0.0) to[out=0,in=270] +(0.65, 0.1);
	\draw (3.25,-0.05) to[out=270,in=180] +(0.1, -0.1) -- +(0.55, 0) to[out=0,in=90] +(0.65, -0.1)
	+(0.0, 0.0) to[out=90,in=180] +(0.1, 0.1) -- +(0.55, 0.0) to[out=0,in=270] +(0.65, 0.1);
	\draw (4.85,-0.05) to[out=270,in=180] +(0.1, -0.1) -- +(0.55, 0) to[out=0,in=90] +(0.65, -0.1)
	+(0.0, 0.0) to[out=90,in=180] +(0.1, 0.1) -- +(0.55, 0.0) to[out=0,in=270] +(0.65, 0.1);
	\draw (6.45,-0.05) to[out=270,in=180] +(0.1, -0.1) -- +(0.55, 0) to[out=0,in=90] +(0.65, -0.1)
	+(0.0, 0.0) to[out=90,in=180] +(0.1, 0.1) -- +(0.55, 0.0) to[out=0,in=270] +(0.65, 0.1);
	
	\tikzset{shift={(0, 0.2)}}
	\draw[gray] (-0.2, -1.25) -- (8.2, -1.25);
	\draw[gray] (-0.2, -1.00) -- (8.2, -1.00);
	\draw[gray] (-0.2, -0.75) -- (8.2, -0.75);
	\draw[gray] (-0.2, -0.5 ) -- (8.2, -0.5 );
	\fill[draw=black,fill=black!10!] (0,-1.5) rectangle (1.6,-0.5);
	\fill[draw=black,fill=black!10!] (1.6,-1.5) rectangle (3.2,-0.5);
	\fill[draw=black,fill=black!10!] (3.2,-1.5) rectangle (4.8,-1.0);
	\fill[draw=black,fill=black!10!] (4.8,-1.5) rectangle (6.4,-1.0);
	\fill[draw=black,fill=black!10!] (6.4,-1.5) rectangle (8.0,-1.0);
	\draw (-0.2, -1.5) -- (8.2, -1.5);
	\draw (0.0, -1.5) -- +(0.0, -0.1);
	\draw (1.6, -1.5) -- +(0.0, -0.1);
	\draw (3.2, -1.5) -- +(0.0, -0.1);
	\draw (4.8, -1.5) -- +(0.0, -0.1);
	\draw (6.4, -1.5) -- +(0.0, -0.1);
	\draw (8.0, -1.5) -- +(0.0, -0.1);
	
	\draw (0.8, -1.75) node {$x_1$};
	\draw (2.4, -1.75) node {$x_2$};
	\draw (4.0, -1.75) node {$x_3$};
	\draw (5.6, -1.75) node {$x_4$};
	\draw (7.2, -1.75) node {$x_5$};
	
	\fill[fill=black!10!] (3.0, -1.45) rectangle node {histogram of packet arrival} +(2.0, 0.4);
	
	\tikzset{shift={(0, -1.8)}}

	\draw[line width=0.3mm] (-0.1,-0.05) to[out=270,in=180] (0.0, -0.15) -- (1.5, -0.15) to[out=0,in=90] (1.6, -0.25)
	+(0.0, 0.0) to[out=90,in=180] +(0.1, 0.1) -- (8, -0.15) to[out=0,in=270] (8.1, -0.05);
	\draw[->,line width=0.3mm] (1.6, -0.25) -- (1.6, -0.5);
	
	\begin{scope}[shift={(-0.4,-0.3)}]
	\begin{scope}[shift={(2.0,-0.2)}]
	
	\draw[line width=0.3mm] (-1.0, -0.6) to[in=90,out=180] (-1.1, -0.7) -- (-1.1, -2.45) to[in=0,out=270] (-1.2, -2.55) node[anchor=east] {5 $\times$} (-1.2, -2.55) to[in=90,out=0] (-1.1, -2.65) -- (-1.1, -4.4) to[out=270,in=180] (-1.0, -4.5);
	
	\draw[->,line width=0.3mm] (0,-0.4) -- +(0,-0.1) to[out=270,in=180] +(0.1, -0.2) -- +(0.9, -0.0) to[out=0,in=90] +(1.0, -0.1) -- +(0, -2.4) to[out=270,in=0] +(-0.1, -2.5) -- +(-0.8, -0);
	
	\draw (-0.9,0.0) rectangle node {64 $\times$ Conv 1} +(1.8,-0.4);
	\draw[->,line width=0.3mm] (-0.0, -0.4) -- +(0, -0.4);	
	\draw (-0.9,-0.8) rectangle node {64 $\times$ Conv 3} +(1.8,-0.4);
	\draw (-0.9,-1.2) rectangle node {BN} +(1.8,-0.4);
	\draw (-0.9,-1.6) rectangle node {ReLU} +(1.8,-0.4);
	\draw (-0.9,-2.0) rectangle node {64 $\times$ Conv 3} +(1.8,-0.4);
	\draw (-0.9,-2.4) rectangle node {BN} +(1.8,-0.4);
	
	\draw[->,line width=0.3mm] (0.0,-2.8) -- +(0.0, -0.2);
	\draw (0.0,-3.2) circle [radius=0.2];
	\draw[line width=0.25mm] (-0.15,-3.2) -- (0.15, -3.2);
	\draw[line width=0.25mm] (0.0,-3.05) -- (0.0, -3.35);
	\draw[->,line width=0.3mm] (0.0,-3.4) -- +(0.0, -0.2);
	
	\draw (-0.9,-3.6) rectangle node {ReLU} +(1.8,-0.4);
	\draw (-0.9,-4.0) rectangle node {MaxPool 4} +(1.8,-0.4);

	\end{scope}
	
	\draw[->,line width=0.3mm] (2.0, -4.6) -- (2.0, -4.7) to[in=180,out=270] (2.1, -4.8) -- (3.5, -4.8)  to[in=270,out=0] (3.6, -4.7) -- (3.6, -1.2)  to[in=180,out=90] (3.7, -1.1) -- (5.9, -1.1)  to[in=90,out=0] (6.0, -1.2) -- (6.0, -1.5);
	
	\begin{scope}[shift={(6.0,-1.5)}]
	
	\draw (-0.9, 0.0) rectangle node {LSTM 100} +(1.8,-0.4);
	\draw[->,line width=0.3mm] (0.9, -0.05) -- (1.0,-0.05) to[in=90,out=0] (1.1, -0.15) to[in=0,out=-90] (1.0, -0.25) -- (0.9, -0.25);
	\draw (-0.9, -0.4) rectangle node {BN} +(1.8,-0.4);
	
	\draw[line width=0.3mm] (-1.0, 0.1) to[in=90,out=180] (-1.1, -0.0) -- (-1.1, -0.3) to[in=0,out=270] (-1.2, -0.4) node[anchor=east] {3 $\times$} (-1.2, -0.4) to[in=90,out=0] (-1.1, -0.5) -- (-1.1, -0.8) to[out=270,in=180] (-1.0, -0.9);
	
	\draw[->,line width=0.3mm] (0.0,-0.8) -- +(0.0,-0.2) to[in=180,out=270] +(0.1,-0.3) -- +(0.8,-0.0) to[in=90,out=0] +(0.9,-0.1) -- +(0.0,-0.1);
	
	\draw[->,line width=0.3mm] (0.0,-0.8) -- +(0.0,-0.2) to[in=0,out=270] +(-0.1,-0.3) -- +(-0.8,-0.0) to[in=90,out=180] +(-0.9,-0.1) -- +(0.0,-0.1);
	
	\draw (-1.9,-1.3) rectangle node {3 $\times$ Conv 1} +(1.8,-0.4);
	\draw (-1.9,-1.7) rectangle node {Softmax} +(1.8,-0.4);
	\draw (-1.0,-2.3) node {variant} +(1.2,-0.4);
	
	\draw (0.1,-1.3) rectangle node {2 $\times$ Conv 1} +(1.8,-0.4);
	\draw (0.1,-1.7) rectangle node {Softmax} +(1.8,-0.4);
	\draw (1.0,-2.3) node {pacing};
	
	\end{scope}
	\end{scope}
	\end{tikzpicture}
 	\caption{Architecture of \name{}. Captured packets are at first assigned to histogram bins with regard to arrival time. The packet arrival histogram is then used in a 1D-CNN with added unidirectional LSTM RNN layers and a final splitted classification layer.}
	\label{fig:basicarch}
\end{figure}

We now present the architecture of \name{} along \fig{fig:basicarch}.

\afblock{CC Manifestation in Traffic.}
We use passive traffic captures to identify the CC variant.
Our only input is the packet arrival time of a flow, we thus only assume packet timing information (\eg{} unlike Netflow) and that we are able to associate packets to flows.
Since any CC variant will effectively control the sending of packets in terms of how many will be sent and at which point in time, we argue that packet timing data inherently captures a CC's behavior.
We associate the packet arrivals into constant-sized bins to get a histogram of packet arrivals with equidistant timesteps for a fixed timing structure. We denote this histogram as $X=[x_0, ..., x_t]$ where every $x_i$ is a one-dimensional feature.
The timing structure is then exploited by the convolutions of our neural network.

\afblock{Packet Arrival Times as the Only Feature.}
We feed the histogram of packet arrivals $X$ into a deep neural network (DNN).
Our primary motivation for using deep learning is its ability to learn features from data~\cite{LeCun:CNN}.
Therefore, we do not engineer any features other than packet timing---the signature of every CC---to ensure the versatility of our approach.

\afblock{Deep Learning-based Classification.}
Inspired by convolutional, long short-term memory DNN (CLDNN) approaches used in speech recognition~\cite{sainath:cldnn}, our DNN consists of a convolutional neural network (CNN)~\cite{LeCun:CNN} and a long short-term memory (LSTM)~\cite{Hochreiter:LSTM} part (see lower part of \fig{fig:basicarch}).
First, the histogram is processed by a CNN which we regard as a feature extraction stage and is derived from a 2D VGGNet-13~\cite{Simonyan:VGG} which is mainly used for image recognition.
We use 5 subsequent modules, which consist of 1D convolutional layers, ReLU activations and 1D maxpooling layers.
We combine two 1D convolutional layers with 64 filters of size 3 and stride 1 with a ReLU activation in between.
We then add the input to the second convolutional layer output as identity skip connection and add a final ReLU activation. To be able to add the input to the output w.r.t. the dimensions, we add a first convolution with 64 filters of filtersize 1 to increase the dimensionality of the histogram input.
We adopted the skip connections from the residual learning~\cite{He:Resnet} approach, as it allowed our network to train more easily and achieve better inference results.
In contrast to residual networks, we use a maxpooling layer instead of convolutions with higher strides to reduce the dimensionality after each module, as our convolutions of size 3 are smaller than the pooling strides of 4 we are using.
Moreover, we use a batch normalization~\cite{Ioffe:Batchnorm} layer after every convolution to normalize the activations in turn reducing overfitting and improving training speed.
The CNN part of the network reduces the incoming data dimensionality by a factor of 1024.
1024 timesteps (\eg{} 1024ms with 1ms binsize) are reduced into a single LSTM timestep.
This timestep is then used in three subsequent unidirectional LSTM~\cite{Hochreiter:LSTM} layers with 100 memory units.
If only fixed sequences of packets should be identified, it proved to be sufficient to also use a fully connected layer as a standard VGGNet does.
However, as we are interested in identifying varying length traffic flows, we use LSTM recurrent neural network (RNN) layers to build up a memory depending on previous behavior.
Following every LSTM layer, we again incorporate a batch normalization layer for the same reasons as before.
After the last LSTM layer, we compute 3 + 2 logits using convolutional layers with filtersize 1.
We split the logits for two different softmax classifications: 3 logits are used for CC classification (\eg{} RENO, CUBIC, BBR), while
2 logits are used for classification whether pacing was used.
This allows us to split the loss function into two crossentropy-losses which we can weight accordingly to emphasize correct CC variant rather than correct pacing classification during training.

\section{Evaluation}
\label{sec:eval}
We next evaluate \name{} in a testbed setup by focusing on CUBIC, RENO, and BBR as major CC variants.
We first describe our setup used for training data generation and testing.
Subsequently, we present how well \name{} identifies CC variants and put a special focus on where it struggles.

\subsection{Experimental Setup}
Since \name{} bases on supervised deep learning, we need labeled data of traffic flows in order to train the neural network.
Labeled data of traffic flows with respect to CC is scarce, so we generate this data on an experimental testbed.

\afblock{Mininet-based Network Testbed.}
We utilize Mininet~\cite{Lantz:mininet} to generate traffic in two topologies subject to different network conditions.
We vary the number of TCP senders, the link latency, the bottleneck link bandwidths, and its queue sizes defined as multiples of the bottleneck link's BDP.
The senders send 60s of fully-loaded TCP streams using a chosen CC variant (in this paper, we focus on BBR, CUBIC, RENO - when using pacing we note ``-p`` to the name) from a standard Linux 4.18 kernel.
Other senders than the to be observed sender (if available) start 2s prior.
In each setting, we capture traffic before and after the bottleneck link in the respective network that is used to train and evaluate our approach.

\begin{table}
	\centering
	\begin{tabularx}{\columnwidth}{|l|X|}
		\hline
		\textbf{Bandwidth} (\mbps)   & 1, 2,..,10, 20, 25, 30, 40, ..., 100, 150, 200                                              \\ \hline
		\textbf{Latency} (ms)       & 1, 2, ..., 20, 30, ..., 100, 150, ..., 300 \\ \hline
		\textbf{BDP Factor}         & 0.5, 1, 2, 5, 10                                                                                                          \\ \hline
	\end{tabularx}
	\caption{Parameters for the single-host network}
	\vspace{-1.5em}
	\label{tab:netparamssingle}
\end{table}

\begin{table}
	\centering
	\begin{tabularx}{\columnwidth}{|l|X|}
		\hline
		\textbf{Bandwidth} (\mbps)   & 1, 2, 3, 4, 5, 10, 15, 20, 25, 30, 40, 50, 100 \\ \hline
		\textbf{Latency} (ms)       & 1, 2, 5, 10, 20, 50 \\ \hline
		\textbf{BDP Factor}         & 1, 5, 10             \\ \hline
	\end{tabularx}
	\caption{Parameters for the multi-host network.}
	\vspace{-1em}
	\label{tab:netparamsthree}
\end{table}

\begin{figure*}
	\centering
	\hspace*{0.42in}
	\includegraphics{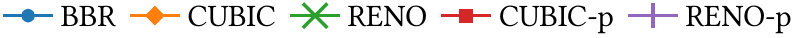}
	\\
	\begin{subfigure}[t]{2.02in}
		\centering
		\includegraphics{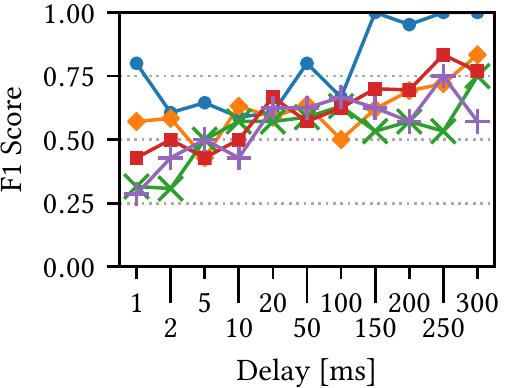}
		\vspace{-1.7em}
		\captionsetup{justification=raggedleft,singlelinecheck=false,margin=21pt}
		\caption{Single-Host 2\mbps{}}
		\label{fig:f1-singlehost-2mbps}
	\end{subfigure}%
	\begin{subfigure}[t]{1.6in}
		\centering
		\includegraphics{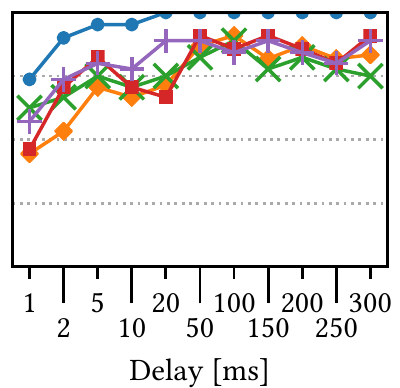}
		\vspace{-1.7em}
		\caption{Single-Host 10\mbps{}}
		\label{fig:f1-singlehost-10mbps}
	\end{subfigure}%
	\begin{subfigure}[t]{1.6in}
		\centering
		\includegraphics{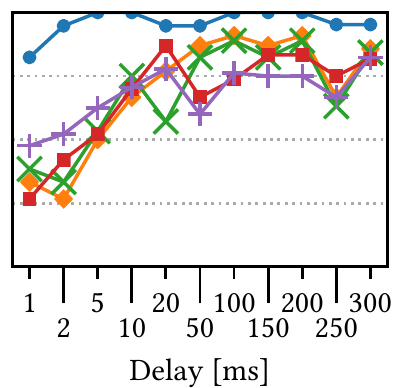}
		\vspace{-1.7em}
		\caption{Single-Host 25\mbps{}}
		\label{fig:f1-singlehost-30mbps}
	\end{subfigure}%
	\begin{subfigure}[t]{1.6in}
		\centering
		\includegraphics{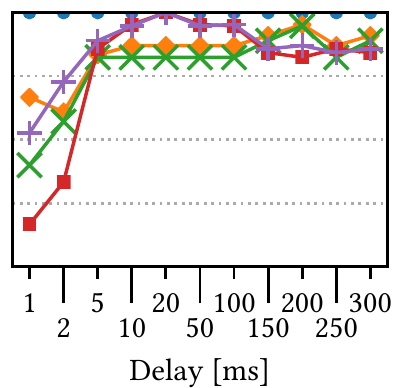}
		\vspace{-1.7em}
		\caption{Single-Host 50\mbps{}}
		\label{fig:f1-singlehost-50mbps}
	\end{subfigure}%
    \\
   	\vspace{0.5em}
	\begin{subfigure}[t]{2.02in}
		\centering
		\includegraphics{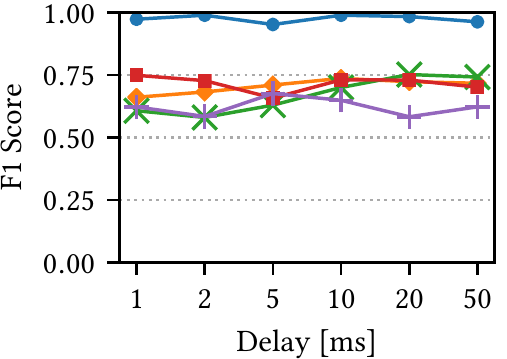}
		\captionsetup{justification=raggedleft,singlelinecheck=false,margin=20pt}
		\vspace{-1.7em}
		\caption{Multi-Host 2\mbps{}}
		\label{fig:f1-multihost-2mbps}
	\end{subfigure}%
	\begin{subfigure}[t]{1.6in}
		\centering
		\includegraphics{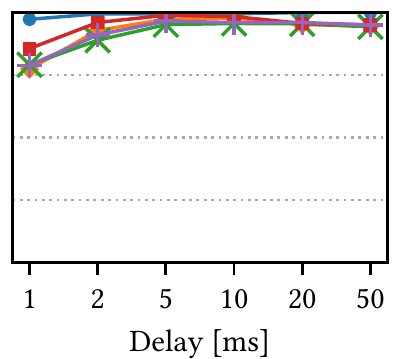}
		\vspace{-1.7em}
		\caption{Multi-Host 10\mbps{}}
		\label{fig:f1-multihost-10mbps}
	\end{subfigure}%
	\begin{subfigure}[t]{1.6in}
		\centering
		\includegraphics{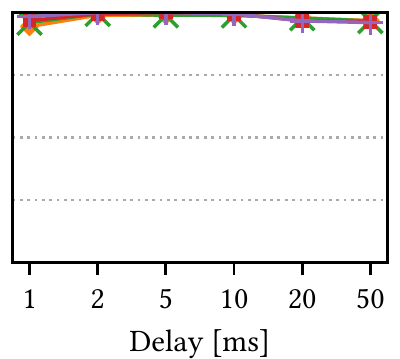}
		\vspace{-1.7em}
		\caption{Multi-Host 25\mbps{}}
		\label{fig:f1-multihost-30mbps}
	\end{subfigure}%
	\begin{subfigure}[t]{1.6in}
		\centering
		\includegraphics{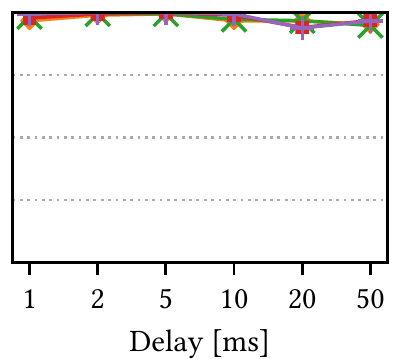}
		\vspace{-1.7em}
		\caption{Multi-Host 50\mbps{}}
		\label{fig:f1-multihost-50mbps}
	\end{subfigure}%
   	\vspace{-0.5em}
	\caption{F1 scores for single and multi-host network \wrt{} bandwidth and delay.}
	\label{fig:bw-delay}
\end{figure*}

\afblock{Single-Host Network.}
The single-host network serves as a baseline condition where only a single TCP sender is active---the simplest condition for CC detection.
It consists of a dumbbell topology where the sending host is connected to a router which is connected via a bottleneck link to a router which is connected to another host.
No background traffic is generated.
The specific parameters used in this network topology are shown in Table~\ref{tab:netparamssingle}.

\afblock{Multi-Host Network.}
More complex than the single-host network, here, three hosts reside on each side of the network.
The hosts use all possible combinations of the CC variants such that we generate traffic with impact of all other CC variants.
We reduced the parameter set to counter the growing parameter space for all combinations.
The specific parameters used are shown in Table~\ref{tab:netparamsthree}.

\afblock{Cross-Traffic Network.}
In the cross-traffic network, we emulate the influence of 3 side flows which cross 4 main flows on parts of a route.
The side flows traverse a 25\mbps{}, a shared 50\mbps{} and another 25\mbps{} link.
The main flows traverse 3 links limited to 50\mbps{}.
The second link is shared with the side flows in the same sending direction.
The latency of all links is fixed to 10ms, while the queue sizes can be either 1, 5 or 10 times the BDP.
The side flows and 3 main flows use CUBIC, RENO, and BBR as CC to emulate background traffic.
The fourth main flow uses a CC of choice.
We capture traffic before and after the shared link.
Thus, this network shows the influence of vantage points where not all traffic is visible and where CCs with different bottlenecks interact.

\afblock{Training.}
As training data, we use the single and multi-host network datasets with vantage points before and after the bottleneck.
We utilize bottleneck link rates of 4\mbps{} and 30\mbps{} as the validation set on which we optimized the DNN and 2\mbps{}, 10\mbps{}, 25\mbps{} and 50\mbps{} bottlenecks as the test set.
This fixed holdout allows us to use the same trained model during all of the evaluation and to reason more in detail about certain results.

We use 1ms as histogram bin-size such that the histogram contains $60000$ entries after 60s of traffic.
For training, we utilize backpropagation through time unrolling all LSTM timesteps.
Moreover, we train the model with the Adam\cite{Kingma:Adam} optimizer with a learning rate of $0.001$ and a learning rate schedule to halve it every 5 epochs, use a batch size of 32 and early stop after 5 epochs of no validation loss improvement. As loss, we use the 1:4 weighted cross-entropy between pacing and variant classification averaged over all timesteps and batch entries (to put more emphasis on the variant).

\subsection{\name{} Performance}

\subsubsection{Identification Performance by Delay and Bandwidth}
\label{sec:eval1}
At first, we evaluate the identification performance \wrt{} delay and bandwidth.
For this, we evaluate the F1 scores of all variants per bandwidth and delay for the single and the multi-host network before and after the bottleneck.
The F1 scores are shown in Figure \ref{fig:bw-delay}.

We can see that our approach distinguishes very well between BBR and loss-based congestion for larger bandwidths above 10\mbps{}.
Bandwidths $\geq$ 10\mbps{} and delays $\geq$ 5ms lead to individual F1 scores above 90\% in the more realistic multi-host scenarios and F1 scores above 55\% in the single-host scenario while BBR is identified with F1 scores above 90\%.
Including also the smaller bandwidths, the minimum F1 score drops onto 40\% in the single-host case and 55\% in multi-host case. Including also the smaller delays decreases the performance for the single-host network further with F1 scores below 20\%.
In general, we can see that larger bandwidths, larger delays and multiple competing flows are beneficial for our approach.

We attribute the effect of the bandwidth to the integer discretization of the congestion window.
The maximum cwnd depends on the bottleneck bandwidth.
Therefore, more steps of, \eg{} the cubic behavior of CUBIC are sampled with larger bandwidths.
Lower bandwidths imply fewer sampled steps and the cubic behavior is harder to discriminate against, \eg{} linear behavior as with RENO.

Next, we could see that higher delays increased the identification performance.
We attribute the effect of the delay to our histogram bin-size.
If the delay is too small, the change in packet arrival is too fast for the time sub-sampling of the bins such that the behavior cannot be derived if it does not differ drastically enough (e.g. with higher bandwidths).
Moreover, the delay also affects the decision of whether pacing was used due to too many packets (being paced or not) falling into the same bin (not shown).

However, we can see that also the multi-host case achieves better results for the same delays. We attribute this effect to the competition of the flows. Beside filling the queues, increasing the queuing delays and leading to more congestion, these hosts also imply a competition for packets in the queue. While the rate of a single-host flow quickly does not increase with an increasing cwnd when the link is saturated, increasing the flow's cwnd in the multi-host scenario can increase the share of its packets in the queue in turn increasing its rate. The individual changes to the cwnd of the different congestion control variants therefore have a higher influence onto the rate which affects the packet-arrival stronger and over a longer time contrasting the issues of smaller delays and bandwidths.

As we have seen, bandwidth and delay impact our approach and too small delays / bandwidths result in low identification performance.
For our further evaluation we will continue with 50\mbps{} as bandwidth to better see where the approach is further challenged.

\subsubsection{Vantage Point before Bottleneck.}
We now evaluate the identification performance of our model with only a vantage point \textit{before} the network bottleneck.
Here, we see packets \textit{before} they are dropped due to congestion or shaped due to queues and thus capture packet arrival times as generated by CC.
We, therefore, expect very good identification results.

\afblock{Single-Host Network.}
The single-host identification performance before the network can be seen in Table~\ref{tab:single-host-edge} (values not in brackets). The upper results in a cell were evaluated on the same delays as for the multi-host scenario (mh), while the lower results use all available delays in the test-set (all). We employ this split as our single-host test set contains more delays but we would like to maintain comparability.
As expected, the results are good with an F1 macro average and overall accuracy of 89\% when observing the same delays as in the multi-host network. With all delays, the accuracy increases to 98\% due to the inclusion of higher delays.

\afblock{Multi-Host Network.}
The multi-host (\ie{} when having competing traffic of other hosts) identification performance is shown in Table~\ref{tab:multi-host-edge} (values not in brackets).
The results are significantly better as already observed before with an overall accuracy of 99\% and a macro F1 average of 99\%.
Comparing the single and multi-host performance, we can see a significant drop in recall for CUBIC-p and decreased precision for CUBIC in the single-host scenario. Moreover, also precision and recall are smaller for RENO and RENO-p.
Indeed, CUBIC-p flows are often erroneously classified as CUBIC and RENO flows are confused with RENO-p and vice versa (not shown due to space limits). This also explains why higher delays are beneficial, as this way the pacing is easier to distinguish.

\subsubsection{Vantage Point after Bottleneck.}
When applied to realistic Internet traffic, CC identification must be robust to packet arrival patterns being altered by queues and other elements.
We therefore now evaluate the performance on traffic \textit{after} being shaped by the bottleneck link on the same trained model as before.

\begin{table}
	\centering
	\setlength\tabcolsep{0.39em}
\begin{tabular}{rr|c|c|c|c|}
	\cline{3-6}
	                                              &  & Precision & Recall & F1 & Accuracy \\ \hline
	\multicolumn{1}{|r}{\multirow{2}{*}{BBR}}     & mh  & 1.00 (1.00) & 1.00 (1.00) & 1.00 (1.00) & 1.00 (1.00) \\
	\multicolumn{1}{|r}{}                         & all & 1.00 (0.99) & 1.00 (1.00) & 1.00 (1.00) & 1.00 (1.00) \\ \hline
	\multicolumn{1}{|r}{\multirow{2}{*}{CUBIC}}   & mh  & 0.78 (0.59) & 0.97 (0.90) & 0.87 (0.71) & 0.94 (0.85) \\
	\multicolumn{1}{|r}{}                         & all & 0.95 (0.63) & 0.99 (0.92) & 0.97 (0.75) & 0.99 (0.88) \\ \hline
	\multicolumn{1}{|r}{\multirow{2}{*}{RENO}}    & mh  & 0.90 (0.77) & 0.87 (0.33) & 0.88 (0.47) & 0.95 (0.85) \\
	\multicolumn{1}{|r}{}                         & all & 0.98 (0.89) & 0.97 (0.47) & 0.98 (0.61) & 0.99 (0.88) \\ \hline
	\multicolumn{1}{|r}{\multirow{2}{*}{CUBIC-p}} & mh  & 0.96 (0.83) & 0.73 (0.67) & 0.83 (0.74) & 0.94 (0.91) \\
	\multicolumn{1}{|r}{}                         & all & 0.99 (0.91) & 0.93 (0.81) & 0.96 (0.85) & 0.98 (0.94) \\ \hline
	\multicolumn{1}{|r}{\multirow{2}{*}{RENO-p}}  & mh  & 0.87 (0.85) & 0.90 (0.73) & 0.89 (0.79) & 0.95 (0.92) \\
	\multicolumn{1}{|r}{}                         & all & 0.96 (0.90) & 0.98 (0.88) & 0.97 (0.89) & 0.99 (0.95) \\ \hline
	\multicolumn{1}{|r}{}                          & mh  & \multicolumn{4}{c|}{\multirow{2}{*}{Overall Accuracy:~\begin{tabular}{@{}c@{}}0.89 (0.73)\\0.98 (0.81)\end{tabular}}} \\
	 \multicolumn{1}{|r}{}                        & all & \multicolumn{4}{c|}{} \\ \hline
\end{tabular} 	\caption{Identification metrics for single-host network before (after) 50\mbps{} bottleneck for all delays or delays as in the multi-host network (mh) for comparability.}
	\vspace{-2em}
	\label{tab:single-host-edge}
\end{table}

\begin{figure}
	\centering
	\includegraphics{gen/delay-legend}
	\\
	\vspace{5pt}
	\begin{subfigure}[t]{1.86in}
		\centering
		\includegraphics{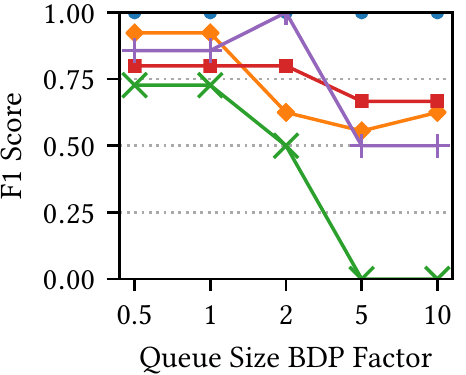}
		\captionsetup{justification=raggedleft,singlelinecheck=false,margin=26pt}
		\caption{Single-Host}
		\label{fig:f1-bdp-single-host-after}
	\end{subfigure}%
	\hfill
	\begin{subfigure}[t]{1.47in}
		\centering
		\includegraphics{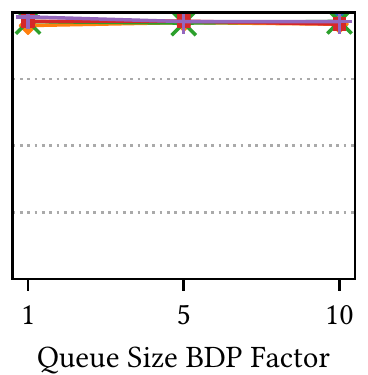}
		\caption{Multi-Host}
		\label{fig:f1-bdp-multi-host-after}
	\end{subfigure}%

	\caption{F1 scores for single and multi-host network after a 50\mbps{} bottleneck \wrt{} queue size for the same delays as in the multi-host network for comparability.}
	\label{fig:f1-bdp-host-after}
\end{figure}

\afblock{Single-Host Network.}
For the single-host network, we can see some interesting differences.
The identification performance is again shown in Table~\ref{tab:single-host-edge} (values in brackets).
Our approach is more challenged with data after the bottleneck which can be easily explained by the queuing dynamics, smoothing the behavior, making it harder to identify certain characteristics of CC.
While most recalls decrease slightly, RENO's recall decreases to 33\%.
We found a correlation between the queue size (as a factor of the BDP) and RENO's F1 score, which is shown in Figure~\ref{fig:f1-bdp-single-host-after}.
For BDP factors $0.5$ and $1$, RENO is classified with F1 scores above 70\% but for greater factors, it reduces to 50\% for twice of the BDP and 0\% for 5 and 10 times the BDP.
In fact, many of the RENO flows are classified as CUBIC (not shown), which also results in worse precision and hence worse F1 score for CUBIC.

Our approach, therefore, seems to be prone to bufferbloat~\cite{Gettys:bufferbloat} \wrt{} RENO.
We account this effect to the shaping nature of queues, which in essence paces out data smoothly when the link is saturated causing RENO as well as CUBIC to look more similar.
This also explains why paced RENO can be distinguished better: Packets are more equally distributed over an RTT and there is an increased chance that the queue is already drained if the next packet arrives, so the queue size is less impacting the smoothing.

\begin{table}
	\centering
	\setlength\tabcolsep{0.39em}
\begin{tabular}{rc|c|c|c|c|}
	\cline{2-5}
	 & \multicolumn{1}{|c|}{Precision} & Recall & F1 & Accuracy \\ \hline
	\multicolumn{1}{|r}{BBR} & \multicolumn{1}{|c|}{1.00 (1.00)} & 1.00 (1.00) & 1.00 (1.00) & 1.00 (1.00) \\ \hline
	\multicolumn{1}{|r}{CUBIC} & \multicolumn{1}{|c|}{0.99 (0.94)} & 0.99 (0.97) & 0.99 (0.96) & 1.00 (0.98) \\ \hline
	\multicolumn{1}{|r}{RENO} & \multicolumn{1}{|c|}{0.99 (0.98)} & 0.99 (0.95) & 0.99 (0.96) & 1.00 (0.99) \\ \hline
	\multicolumn{1}{|r}{CUBIC-p} & \multicolumn{1}{|c|}{0.99 (0.95)} & 1.00 (0.97) & 0.99 (0.96) & 1.00 (0.99) \\ \hline
	\multicolumn{1}{|r}{RENO-p} & \multicolumn{1}{|c|}{1.00 (0.98)} & 0.99 (0.96) & 0.99 (0.97) & 1.00 (0.99) \\ \hline
	\multicolumn{1}{r}{} & \multicolumn{4}{|c|}{Overall Accuracy: 0.99 (0.97)} \\ \cline{2-5}
\end{tabular} 	\caption{Identification metrics for multi-host network before (after) 50\mbps{} bottleneck.}
	\vspace{-1.5em}
	\label{tab:multi-host-edge}
\end{table}

\afblock{Multi-Host Network.}
When evaluating the multi-host network, we see more expected results.
The identification performance is again shown in Table~\ref{tab:multi-host-edge} (values in brackets).
The overall accuracy reduces only insignificantly by 2 percent-points onto 97\%.
We are therefore convinced that our approach's susceptibility to bufferbloat alleviates with concurrent flows. 
We attribute this to the stronger influence of the cwnd onto the flow rate when multiple flows compete as we described before in \sref{sec:eval1}.

\begin{table}
	\centering
	\setlength\tabcolsep{0.39em}
\begin{tabular}{rc|c|c|c|c|}
	\cline{2-5}
	 & \multicolumn{1}{|c|}{Precision} & Recall & F1 & Accuracy \\ \hline
	\multicolumn{1}{|r}{BBR} & \multicolumn{1}{|c|}{1.00 (1.00)} & 1.00 (1.00) & 1.00 (1.00) & 1.00 (1.00) \\ \hline
	\multicolumn{1}{|r}{CUBIC} & \multicolumn{1}{|c|}{0.98 (0.97)} & 0.88 (0.92) & 0.93 (0.94) & 0.97 (0.98) \\ \hline
	\multicolumn{1}{|r}{RENO} & \multicolumn{1}{|c|}{0.89 (0.92)} & 0.98 (0.97) & 0.93 (0.95) & 0.97 (0.98) \\ \hline
	\multicolumn{1}{|r}{CUBIC-p} & \multicolumn{1}{|c|}{0.97 (0.96)} & 0.93 (0.89) & 0.95 (0.93) & 0.98 (0.97) \\ \hline
	\multicolumn{1}{|r}{RENO-p} & \multicolumn{1}{|c|}{0.93 (0.90)} & 0.97 (0.96) & 0.95 (0.93) & 0.98 (0.97) \\ \hline
	\multicolumn{1}{r}{} & \multicolumn{4}{|c|}{Overall Accuracy: 0.95 (0.95)} \\ \cline{2-5}
\end{tabular} 	\caption{Identification metrics for cross-traffic network before (after) shared link}
	\label{tab:crosstraffic}
\end{table}

\subsubsection{Arbitrary Vantage Points.}
Until now, we have investigated what happens if packet arrivals are captured at the edge - either before or after a single link.
However, in reality, a flow will traverse multiple links, so there are multiple vantage points to observe the flow.
Moreover, there might cross flows which have been shaped by different bottlenecks interacting with the flow to be observed.

Thus, we now focus on what happens if we measure at a vantage point in the middle of a network, where flows with different bottlenecks compete.
To evaluate this scenario, we use our initially trained model on the cross-traffic network.

The performance results can be seen in Table~\ref{tab:crosstraffic} and are surprisingly similar to the multi-host network after the bottleneck.
We achieve an overall accuracy of 95\% which indicates that our approach generalizes also to arbitrary vantage points although not trained for with remarkable performance regarding distinguishing between BBR and loss-based CC with an F1 score of 100\%.

\subsubsection{Required Flow Duration.}
\begin{figure}
	\centering
	\includegraphics{gen/delay-legend}\\
	\vspace{5pt}
		\begin{subfigure}[t]{3.34in}
		\centering
		\includegraphics{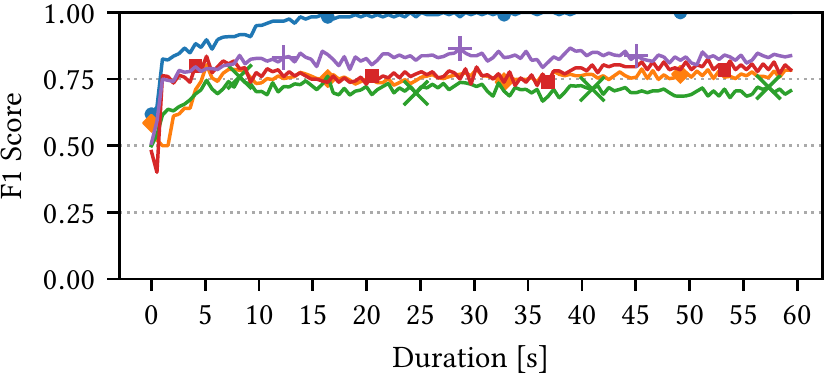}
		\vspace{-1.7em}
		\captionsetup{justification=raggedleft,singlelinecheck=false,margin=80pt}
		\caption{Single-Host}
		\label{fig:f1-singlehost-flow-time}
	\end{subfigure}%
    \\
   	\vspace{5pt}
	\begin{subfigure}[t]{3.34in}
		\centering
		\includegraphics{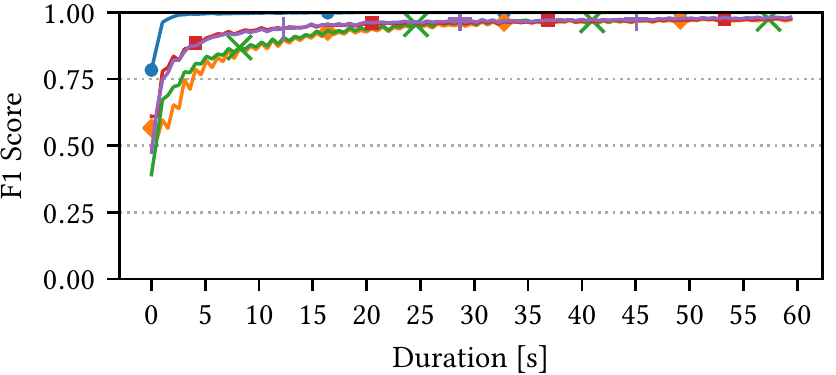}
		\vspace{-1.7em}
		\captionsetup{justification=raggedleft,singlelinecheck=false,margin=81pt}
		\caption{Multi-Host}
		\label{fig:f1-multihost-flow-time}
	\end{subfigure}%
	\vspace{-1em}
	\caption{F1 score \wrt{} flow duration for single and multi-host network. The duration is increased in 512ms steps. The bottleneck was limited to 50\mbps{} and the delays of the multi-host network are used for comparability.}
	\label{fig:f1-flow-time}
	\vspace{-1em}
\end{figure}

The previous results were obtained with flows of 60 seconds length but flows this long occur relatively seldom~\cite{Zhang:flowcharacteristics}.
Therefore, we now evaluate the flow duration needed for correct classification
over the single-host and multi-host network vantage points at 50\mbps{}.
The F1 scores depending on the flow duration sampled in 512ms steps are shown in Figure~\ref{fig:f1-flow-time}.
Generally speaking, we can see that the identification performance increases when flows can be observed over a longer time.

BBR (due to its characteristic behavior in and after slowstart) can be identified with an F1 score of 85\% in 1024ms in the multi-host setting / in 3584ms in the single-host setting.
The unpaced variants need 4096ms to achieve an F1 score over 75\%, while the paced variants need 2048ms to achieve the same in the multi-host setting.
In the single-host setting, 75\% are achieved in 8706ms for the unpaced variants and 1536ms for the paced variants. However, the scores are very unsteady and fluctuate such that, e.g., RENO's F1 score approaches 75\% only for certain moments.

We also found a trivial correlation for loss-based CC between delay and classification duration (longer delay $\hat{=}$ longer duration).
\section{Conclusion}
\label{sec:conclusion}

In this paper, we presented the design and performance of \name{}, a passive method to identify congestion control variants using just packet arrival time information.
\name{} does not need header information, visibility into the reverse path nor domain knowledge during training of the variants that it wants to identify.
It is thus designed for a QUIC-enabled world and can be easily adapted for new congestion control variants.
Our results indicate a promising performance over a diverse set of vantage points.
Especially variants which eminently vary in behavior in comparison to traditional approaches, such as BBR, can be distinguished with remarkable performance throughout most tested scenarios.
Yet, the approach is challenged in distinguishing loss-based CC when it cannot find enough characteristic features, e.g., when only one flow sends data.

We believe that this performance can be partly increased when training and optimizing for specific network properties, \ie{} when knowing where \name{} should be used and how many flows compete, by, e.g., adapting the histogram bin-size.
In the future, we plan to evaluate \name{} on real-world network traces, on further congestion control variants, when subject to AQMs affecting the identification abilities as well as how receiver ACK generation or ACK compression affect the forward path and thus \name.

\begin{acks}
We would like to thank Stefan Koltermann for initially providing us with GPUs.
Further computing resources were granted by RWTH Aachen University under project thes0601.
This work has been funded by the DFG as part of the CRC 1053 MAKI and within the Cluster of Excellence ``Internet of Production'' (IoP) under project ID 390621612. 
\end{acks}

\bibliographystyle{ACM-Reference-Format}
\balance
\bibliography{reference}

\end{document}